\documentclass[fleqn,usenatbib]{mnras}

\usepackage[T1]{fontenc}
\usepackage{hyperref}

\DeclareRobustCommand{\VAN}[3]{#2}
\let\VANthebibliography\thebibliography
\def\thebibliography{\DeclareRobustCommand{\VAN}[3]{##3}\VANthebibliography}

\usepackage{graphicx}	
\usepackage{amsmath}	
\usepackage{amssymb}	
\usepackage{romannum}
\newcommand{\code}{\texttt}

\title[\code{FNet II}: Spectral Classification of Quasars, Galaxies, Stars, and broad absorption line (BAL) Quasars]{\code{FNet II}: Spectral Classification of Quasars, Galaxies, Stars, and broad absorption line (BAL) Quasars}

\author[R.~Moradi et al.,]{
R.~Moradi,$^{1}$\thanks{rmoradi@ihep.ac.cn}
F.~Rastegarnia,$^{2,3,5 }$\thanks{fatemeh.rastegarnia@icranet.org}
Y.~Wang$^{2,3,4}$\thanks{ yu.wang@uniroma1.it}
M.~T.~Mirtorabi,$^{2,5}$\thanks{torabi@alzahra.ac.ir}
\\
$^{1}$Key Laboratory of Particle Astrophysics, Institute of High Energy Physics, Chinese Academy of Sciences,\\ Beijing 100049, People’s Republic of China\\
$^{2}$
ICRANet, Piazza della Repubblica 10, I-65122 Pescara, Italy\\
$^{3}$
ICRA, Dipartimento di Fisica, Universit\`a  di Roma ``La Sapienza'', Piazzale Aldo Moro 5, I-00185 Roma, Italy\\
$^{4}$
INAF -- Osservatorio Astronomico d'Abruzzo,Via M. Maggini snc, I-64100, Teramo, Italy\\
$^{5}$
 Department of Physics,Faculty of Physics and Chemistry, Alzahra University,Tehran, Iran\\
}

\date{Accepted XXX. Received YYY; in original form ZZZ}

\pubyear{2015}

\begin{document}
\label{firstpage}
\pagerange{\pageref{firstpage}--\pageref{lastpage}}
\maketitle

\begin{abstract}
In this work, we enhance the \code{FNet}, a 1-dimensional convolutional neural network (CNN) with a residual neural network (ResNet) architecture, to perform spectral classification of quasars, galaxies, stars, and broad absorption line (BAL)-quasars in the SDSS-IV catalog from DR17 of eBOSS. Leveraging its convolutional layers and the ResNet structure with different kernel sizes, \code{FNet} autonomously identifies various patterns within the entire sample of spectra.  Since \code{FNet} does not require the intermediate step of identifying specific lines, a simple modification  enabled our current network to classify all SDSS spectra. This modification involves changing the final output layer from a single value (redshift) to multiple values (probabilities of all classes), and accordingly adjusting the loss function from mean squared error (MSE) to cross-entropy. \code{FNet} achieves a completeness of 99.00\% $\pm$ 0.20 for galaxies, 98.50\% $\pm$ 0.30 for quasars, 99.00\% $\pm$ 0.18 for BAL-quasars, and 98.80\% $\pm$ 0.20 for stars. These results are comparable to those obtained using \code{QuasarNET}, a standard CNN employed in the SDSS routine, comprises convolutional layers without the ResNet structure with equal kernel sizes, and is utilized for redshift measurement and classification by identifying seven emission lines. \code{QuasarNET}, in order to overcome the problem of finding a CIV emission line with broad absorption which is slightly more challenging than that of detecting emission lines requires to add BAL CIV line to the list of lines that the network learns to identify. However, this procedure is not necessary in \code{FNet} as it learns the features through a self-learning procedure.
\end{abstract}




\section{Introduction}

Classifying astrophysical objects—such as stars, galaxies, and quasars—based on their spectral and photometric properties is crucial for advancing research and understanding the Universe's origin and evolution \citep{2020ApJ...897L..14Y,2020ARA&A..58...27I,2021ApJ...907L...1W,1993MNRAS.263..168H,2001ApJ...551L..27M,2010AJ....140..546W,2020ApJ...891...69C,2021A&A...649A..75M}. 

Paradoxically, the substantial quantity of data generated in astrophysical surveys renders the manual inspection of each spectrum unfeasible. For instance, the seventeenth data release (DR17) from the Sloan Digital Sky Survey (SDSS) encompasses over 623 TB of data on the Science Archive Server (SAS), containing approximately 1.5 million unique spectra and objects \citep{2022ApJS..259...35A}. The SDSS catalog from DR17 of the main survey extended Baryon Oscillation Spectroscopic Survey (DR17-eBOSS) \citep{2016AJ....152..205H,2020ApJS..250....8L,2022ApJS..259...35A}, from which we derive data for our current research, contains 1.4 million spectra. Therefore, automated methods must be implemented to achieve classification with the same precision as human experts.

As of 2023, approximately ten distinct and intermediate methods have been recognized for the classification and identification of stars, galaxies, supernovae, dwarfs, and quasars \citep[see e.g.,][and references therein]{2023MNRAS.518.5904Y}:

1. Template Matching (TM): Constructs a template database for each class and associates new data with the closest matching template \citep{template_matching}. This method is used in various applications, such as the \code{redrock} pipeline classifier in the Dark Energy Spectroscopic Instrument (DESI) quasar survey \citep{2023AJ....165..144G}. See also \citep{subbarao2002sloan, sdssimagepipeline, Westfall_2019, Zhao_2012, glazebrook1998automatic}.

2. K-Nearest Neighbor (KNN): Determines the label for an object based on the majority labels of its K closest objects in the feature space \citep[see e.g.,][]{ZHANG20072038, DENG2016143, 10.1007/s11222-009-9153-8, sookmee2020globular, guzman2018stellar}.

3. Support Vector Machine (SVM): Determines decision boundaries to distinguish between various data classes \citep[see e.g.,][]{brice2019classification, barrientos2020machine, tao2018automated, 2015RAA....15.1137L, liu2021stellar}.

4. Decision Trees (DT): Utilize leaves to represent classification results and nodes to serve as criteria for distinguishing objects \citep{10.1145/234313.234346, zhao2008comparison}.

5. Ensemble Learning: Connects numerous classifiers to form a robust classifier \citep{10.5555/645528.657623, brice2019classification, 2018MNRAS.481.4194M, 2019RAA....19..111L, 2019AJ....158..188B, baqui2021minijpas}.

6. Artificial Neural Networks (ANN): Mimic the functionality of the human brain to perform machine learning tasks \citep[see e.g.,][]{2019MNRAS.485.2167G, 2020PASP..132d4503Z, 2019MNRAS.483.4774L, 2014PASA...31....1F, jing2020new, 2017MNRAS.465.4311W, Rastegarnia_2022, 2018arXiv180809955B, 2020JCAP...11..015F}.

7. Gaussian Naive Bayes: Manages features characterized by a Gaussian distribution, with the outcome determined by the maximum posterior probability \citep[see e.g.,][]{doi:10.1146/annurev.astro.36.1.369, 2010JApA...31..177L, pruzhinskaya2019anomaly, 2015MNRAS.452.4183H}.

8. Bayesian Logistic Regression (LR): Derives posterior probability distributions from linear regression \citep[see e.g.,][]{2021MNRAS.503..484P, tao2018automated, luo2008automated}.

9. Partial Least Squares Discriminant Analysis (PLS-DA): Addresses data within the same cluster, as opposed to data in different clusters \citep{2011AJ....142..203D, SONG201879}.

10. Ranking-based (RB) Classification: Utilizes graph-based ranking methods. An example is the locally linear embedding classifier, which transforms data from its high-dimensional space to a lower-dimensional space based on the manifold \citep[see e.g.,][]{2016PASP..128c4502D, 2018ApJS..234...31L}.

 A comprehensive study by \cite{2023MNRAS.518.5904Y} on spectra in their observed cosmological frame has shown that convolutional neural networks (CNNs) outperform other methods in classifying observed spectra. Since DR16 \citep{2020ApJS..250....8L}, SDSS has officially adopted \code{QuasarNET} \citep{2018arXiv180809955B}, a CNN used to infer redshifts and identify candidates that are likely to be misclassified by the standard pipeline. However, CNNs can face challenges such as overfitting and prolonged computation times due to their strong feature selection capabilities. For optimal classification of stars, galaxies, and quasars, conducting the analysis in the observed cosmological frame is advisable to minimize redshift impact and gain a comprehensive understanding.

Other methods usually perform worse than human-expert classification \citep{2013AJ....145...10D, paris2017sloan, 2018arXiv180809955B}. In contrast, deep learning (DL) methods, with their ability to recognize patterns and identify spectral features such as emission/absorption lines and spectral breaks, achieve accuracy comparable to visual inspection \citep[see e.g.,][]{2018arXiv180809955B}. For example, \cite{Rastegarnia_2022} showed that principal component analysis (PCA) \citep{glazebrook1998automatic} produces extreme redshift estimations for quasars, which are not consistent with visual inspection or DL-based networks like \code{FNet} \citep{Rastegarnia_2022} and \code{QuasarNet} \citep{2018arXiv180809955B}.

Efficient optimization procedures in DL and CNN-based networks have enabled the solution of previously unseen problems \citep{RevModPhys.91.045002}. The successful application of CNNs in large surveys has made them increasingly applicable to astronomical tasks \citep{2022KosNT..28e..27K, 2018arXiv180809955B, 2023arXiv231104146Z, 10.1093/mnras/stad2913, ball2010data, allen2019deep, 2023ApJ...954..164T}.

In this paper, we enhance the previously implemented network \code{FNet} \citep{Rastegarnia_2022} to classify quasars, galaxies, stars, and BAL quasars in the DR17 eBOSS dataset. \code{FNet} is a CNN with a residual neural network (ResNet) structure, designed to be trained using the observed optical spectra/flux of quasars. Initially, it was created to estimate the redshift of quasars in the SDSS-IV quasar catalog from DR16 eBOSS \citep{2016AJ....152..205H, 2020ApJS..250....8L}. \code{FNet} achieves an accuracy of 97.0\% for the velocity difference in redshift, $|\Delta\nu| < 6000~ \rm km/s$, and 98.0\% for $|\Delta\nu| < 12000~ \rm km/s$.

\code{FNet} provides similar accuracy to \code{QuasarNET} but is applicable to a wider range of SDSS spectra, especially those lacking the clear emission lines exploited by \code{QuasarNET} \citep{Rastegarnia_2022}. In DR16Q, around 8600 spectra were not recognized by \code{QuasarNET} due to the absence of at least two emission lines necessary for its line finder units. After visual inspection, approximately 5000 of these spectra were flagged as quasars, and their visual inspection (VI) redshifts were subsequently reported in the DR16Q catalog. \code{FNet} predicts the redshift of these VI quasars with an accuracy of 87.4\% for $|\Delta\nu| < 6000~ \rm km/s$ (see Fig.~4 in \citealp{Rastegarnia_2022}).

Unlike \code{QuasarNET}, \code{FNet} does not require identifying specific lines, allowing straightforward modification to classify all SDSS spectra. This involved changing the final output layer of \code{FNet} from a single value (redshift) to multiple values (class probabilities) and adapting the loss function from mean squared error (MSE) to cross-entropy. Further details can be found in Sec.~\ref{sec:CNN}.

Both \code{FNet} and \code{QuasarNET} are CNN-based networks but have different designs. \code{QuasarNET} consists of 4 convolutional layers with kernel sizes of 10, following the traditional procedure to identify emission lines as local patterns hidden in spectra using its "line finder" units. \code{FNet}, on the other hand, consists of 24 convolutional layers with a ResNet structure and kernel sizes of 500, 200, and 15, enabling it to find both ``local'' and ``global'' patterns in spectra through a self-learning procedure. This makes \code{FNet} applicable to ambiguous spectra and a larger range of redshifts; further details can be found in Sec.~\ref{sec:results}.

In this work, we utilize the SDSS catalog from DR17 of the main survey eBOSS \citep{2016AJ....152..205H, 2020ApJS..250....8L, 2022ApJS..259...35A}, which comprises 1.4 million observations.

In Sec.~\ref{sec:Data}, we briefly review the basics of quasar spectroscopy, the structure of their optical spectra, and the description of SDSS and eBOSS surveys.

In Sec.~\ref{sec:CNN}, we describe the CNN structure developed for this work.

In Sec.~\ref{sec:results}, we present the results of our training and compare them with similar works in the literature.

Finally, in Sec.~\ref{sec:conclusions}, we present the conclusions of the paper.

\section{Data}\label{sec:Data}

The dataset used in this study is derived from the Sloan Digital Sky Survey IV (SDSS-IV) quasar catalog from Data Release 17 (DR17) of the extended Baryon Oscillation Spectroscopic Survey (eBOSS) \citep{2022ApJS..259...35A}. Notably, there have been no changes in the primary eBOSS survey data since DR16; consequently, both DR16 eBOSS and DR17 eBOSS contain the same number of spectra \citep{2016AJ....152..205H, 2020ApJS..250....8L, 2022ApJS..259...35A}. Data is gathered using 500 fibers on a 2k CCD associated with each spectrograph, with wavelength coverage spanning approximately 361 to 1014 nm \citep{2013AJ....146...32S}.

\begin{figure*}
\centering
\textbf{A}\includegraphics[width=0.45\hsize,clip]{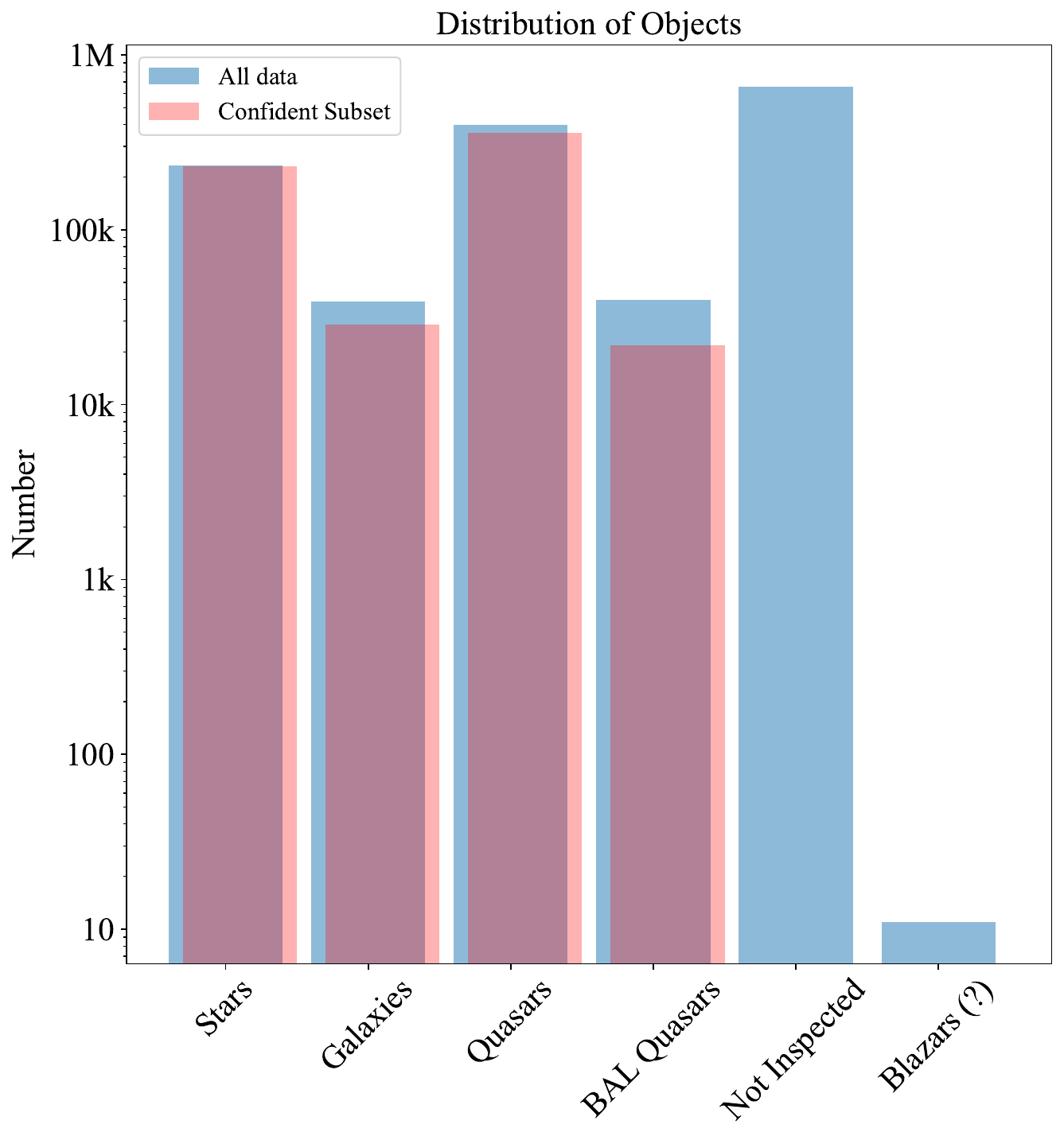}
\textbf{B}\includegraphics[width=0.50\hsize,clip]{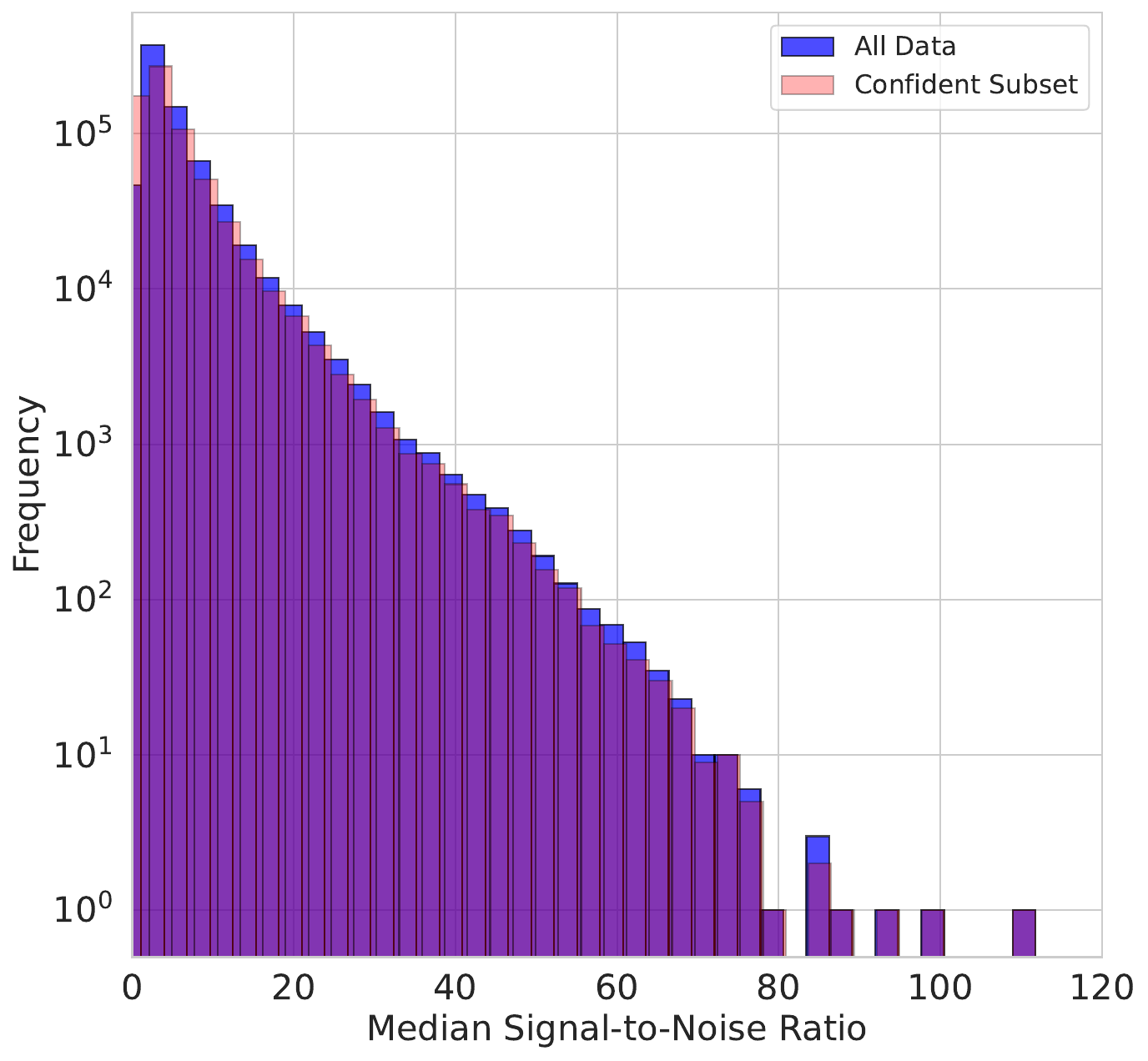}

\caption{\textbf{\textit{Left}}:\textbf{Light blue columns}: The total number of each class in DR17Q is as follows: Stars = 233,859, Galaxies = 39,054, Quasars = 396,843, and BAL Quasars = 39,501. Additionally, 658,020 objects were not inspected, and the remaining {11} objects indicate potential blazars. \textbf{Light red columns}: Based on the criteria introduced in Sec.~\ref{sec:Data} to obtain more confident data for input into \code{FNet}, the total number of each class is as follows: Stars = 229,647, Galaxies = 28,688, Quasars = 359,609, and BAL Quasars = 21,962. {\textbf{\textit{Right}}: Histogram of the median signal-to-noise ratio (SNR). The blue bars represent the SNR for all classes in the DR17Q dataset after excluding ``not inspected'' spectra and potential blazars. The red bars indicate the SNR for the confident subset of the data.}}  
\label{fig:Distribution}
\end{figure*}

The data in this sample is annotated with class identifiers for each spectrum: Star (\code{CLASS\_PERSON} = 1), Galaxy (\code{CLASS\_PERSON} = 4), Quasar (\code{CLASS\_PERSON} = 3), or BAL-Quasar (\code{CLASS\_PERSON} = 30). In total, there are 1,367,288 spectra classified with these identifiers. The distribution is as follows: Stars = 233,859, Galaxies = 39,054, Quasars = 396,843, and BAL-Quasars = 39,501. Additionally, 658,020 objects were designated as \code{CLASS\_PERSON} = 0, indicating they were not inspected. Eleven spectra with high flux and continua matching an archetypal quasar continuum shape, but without recognizable absorption or emission features, have been marked as possible blazars (\code{CLASS\_PERSON} = 50) \citep{2020ApJS..250....8L}. See Fig.~\ref{fig:Distribution}.

A vital aspect of data selection for training neural networks is the preparation of a training/test data sample that maximizes confidence, ensuring the neural network is trained with the most accurate training set possible. The (DR16Q/DR17Q)-eBOSS dataset includes automated classifications determined by version \code{v5\_13\_0} of the SDSS spectroscopic pipeline \citep{2020ApJS..250....8L}. The quasar-only sample is estimated to be contaminated by non-quasar objects at a rate of approximately 0.3\%–1.3\% \citep{2020ApJS..250....8L}. Furthermore, it is affected by low-confidence quasars, stars, and galaxies labeled by a confidence rating recorded in the column \code{Z\_CONF}, with 0 representing the lowest confidence and 3 indicating the highest confidence (a value of -1 indicates the object was not visually inspected). Consequently, due to incorrect pipeline classification and redshift estimation, the DR17Q-eBOSS dataset contains stars and galaxies that have been misclassified as quasars and vice versa \citep{2020ApJS..250....8L, 2021MNRAS.tmp..804F}. To address this, we apply the following criteria to ensure the dataset is as confident as possible:

\begin{enumerate}
    
\item \textbf{High Confidence Level:} Each class is selected based on \code{Z\_CONF} = 3, indicating the highest confidence level.
   
\item \textbf{BAL-Quasar Probability:} To ensure the inclusion of nearly all confident BAL-Quasars, we use \code{BAL\_PROB} $\geq$ 0.9. The \code{BAL\_PROB} represents the probability assigned to each quasar based on the statistical significance of the troughs associated with the C IV line and the quality of the $\chi^2$ fit to the quasar continuum \citep{2020ApJS..250....8L}.
   
\item \textbf{Visual Inspection:} To include only visually inspected sources for Galaxies, Quasars, and BAL-Quasars, we use \code{Z\_VI} > 0.
   
\item \textbf{Quasar Classification:} For quasars, we use \code{IS\_QSO\_FINAL} = 1, while for Galaxies and Stars, we use \code{IS\_QSO\_FINAL} $\neq$ 1. The \code{IS\_QSO\_FINAL} parameter, introduced in DR16, takes integer values from -2 to 2. Quasar spectra have a value of 1, and questionable quasar spectra have a value of 2. All values of 0 or less denote non-quasars \citep{2020ApJS..250....8L}.
\end{enumerate}

{Moreover, in addition to visually inspected quasars from DR7Q \citep{2010AJ....139.2360S} and DR12Q \citep{2017A&A...597A..79P}, some spectra in the dataset were classified by the automated classifier in DR14Q \citep{2018A&A...613A..51P}. Some of these entries were flagged for visual inspection when automated methods failed to recognize their classifications, indicated by the flag \code{AUTOCLASS\_DR14Q = VI}. This subset comprises 35,268 spectra. Many of these spectra exhibit low SNRs or display strong/weak absorption lines, which can confound the pipeline (see Sec.~3 of \cite{2018A&A...613A..51P} and Sec.~3 of \cite{2020ApJS..250....8L}). A small \code{DELTA\_CHI2}, which is the case for many spectra flagged as \code{AUTOCLASS\_DR14Q = VI}, indicates that they are unlikely to be useful for scientific study.}

{Consequently, these spectra of flag \code{AUTOCLASS\_DR14Q = VI} have been excluded from the  dataset since they cannot provide the ground true for training and testing. The final dataset in this study comprises 639,906 spectra, consisting of 229,647 stars, 28,688 galaxies, 359,609 quasars, and 21,962 BAL-quasars (see Fig.~\ref{fig:Distribution} (A).}

{It is important to highlight that the application of the aforementioned criteria does not impose any constraints on the signal-to-noise ratio (SNR) of our test and training sets. The distribution and range of SNR in the confident data used for the training and test sets are representative of the broader dataset, as illustrated in Fig.\ref{fig:Distribution}(B). }

%

\subsection{Prepossessing the data}

To enable deep learning (DL) networks to process data effectively, data preprocessing plays a crucial role. We follow the same procedure presented by \citet{Rastegarnia_2022}. The observed spectrum, representing wavelength versus flux, is akin to a one-dimensional vector. In the DR16Q/DR17Q dataset, each observed spectrum contains nearly 4500 flux data points in the logarithmic space of wavelength. To create trainable data, we fit and extrapolate each spectrum to a one-dimensional row of 4618 data points uniformly distributed in the logarithmic space of wavelength in the range of $360~\rm nm$ to $1032.5~\rm nm$. 

Due to significant variability in the observed flux of the objects, and to accelerate the network and enhance processing accuracy, we normalize the data of each spectrum using zero-mean normalization and unit-norm normalization \citep{jayalakshmi2011statistical}. The reduced, normalized data is then stored in an $N \times M$ matrix, where $N$ represents the total number of objects in different classes and $M = 4618$ is the number of wavelength resolutions. {Specifically, after applying the criteria to ensure a confident dataset, $N = 639,906$.}

\section{Convolutional neural network (CNN)}\label{sec:CNN}

\begin{figure}
\centering
\includegraphics[width=1.0\hsize,clip]{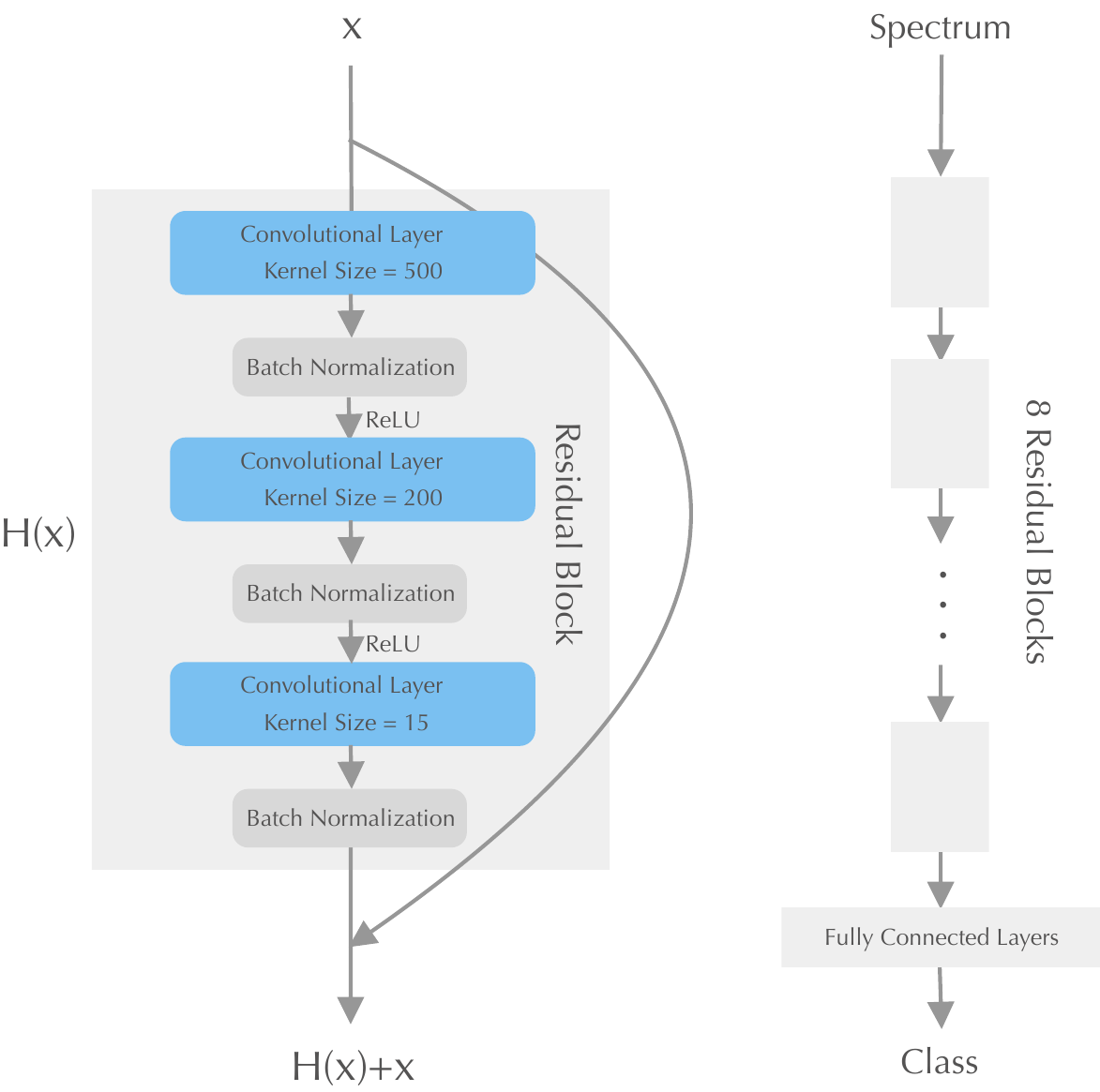}
\caption{The architecture described here, referred to as FNet, uses a 1-dimensional CNN to make classification from the input spectrum. \textbf{Left}: The diagram illustrates the structure of a residual block within the architecture. The input \( x \) is passed through three convolutional layers with kernel sizes of 500, 200, and 15 to detect both global and local patterns in quasar spectrum, resulting in \( H(x) \), and then combined with the original input as \( H(x) + x \). Batch normalization is applied after each convolutional layer, with the ReLU activation function activated after the first two batch normalization layers. \textbf{Right}: The input spectrum goes through 8 residual blocks, the output of these blocks is flattened and processed through fully connected layers, ultimately classifying the spectrum.} 
\label{fig:CNN-structure}
\end{figure}
%

 \subsection{Model Selection}

Convolutional Neural Networks (CNNs) have been widely utilized in machine learning \citep{Goodfellow-et-al-2016}, demonstrating substantial improvements in computer vision \citep{xu2014deep, koziarski2017image, yamashita2018convolutional}, and have become the most widely adopted architecture \citep{yang2015deep, liu2018time}. The architecture of \code{FNet} was initially presented in \citet{Rastegarnia_2022}. In this work, we provide an overview of \code{FNet} and highlight the significant modifications that enable it to excel in classification tasks.

As Residual Networks (ResNets) were employed in the architecture of \code{FNet}, we first examine whether this structure can be enhanced using novel and more complex methods. Introduced by \citet{he2016deep}, ResNets are an advanced form of CNNs that incorporate "skip connections." These improvements allow inputs to bypass certain layers and be added directly to the outputs of later layers, mitigating the vanishing gradient problem and enabling the training of much deeper networks. Currently, ResNet serves as a baseline in many neural network applications and comparisons \citep{wu2017tiny, praveen2022resnet}.

Compared to more complicated or newly introduced architectures, such as Transformers \citep{vaswani2017attention}, ResNets often maintain competitive performance for simpler tasks \citep{Gorishniy2021RevisitingDL}. In our tests of the transformer model with different numbers of headers (2, 4, and 8) and layers (2, 4, and 6), after training with the same data as \code{FNet}, the overall accuracy achieved was always below 95\%, falling short of the performance achieved by the simpler ResNet-based \code{FNet} architecture. It is observed that beyond a certain model complexity, the improvements in accuracy tend to be marginal, especially in scenarios where the data characteristics do not demand such complexity, as is the case for some 1D data \citep{kiranyaz20211d}. Additionally, complexities can sometimes lead to overfitting, particularly when not properly regularized or when the dataset is not sufficiently large \citep{li2019research}. The choice to move to a more complex architecture should be driven by specific needs that cannot be met by simpler models and should be justified by empirical evidence showing clear benefits.

Moreover, the accuracy of any machine learning model is highly dependent on the quality of the input data \citep{whang2020data}. High-quality data are well-curated, balanced, and reflective of real-world scenarios. This includes comprehensive representation of all classes in classification tasks, minimal noise, and high precision. Often, efforts and resources are better spent on improving data quality through enhanced collection methods and rigorous cleaning processes. As shown in Sec.~\ref{sec:Data}, we have carefully preprocessed and selected the data.

Building upon our previous work, where our earlier model achieved high accuracy, as demonstrated by our comparison with the \code{Transformer} and supported by the reasoning above, we continue to utilize \code{FNet} based on ResNet in this article, with some minor modifications for classification instead of regression. Below, we outline the architecture of \code{FNet}.

 \subsection{FNet}

\code{FNet} differs from the officially adopted SDSS \code{QuasarNET} \citep{2018arXiv180809955B}, a CNN-based model that mimics the traditional human method by identifying prominent spectral lines, classifying the spectrum, and calculating redshifts from these lines. In contrast, \code{FNet} is designed to autonomously uncover hidden patterns in the data without relying on predefined information. These patterns could include prominent and weak lines, global patterns, and specific patterns of different classes. This distinct design results in different outcomes and application scopes.

The architecture of \code{FNet} is demonstrated in Fig.~\ref{fig:CNN-structure}. \code{FNet} employs a one-dimensional ResNet architecture designed for classifying spectral data. This architecture features 8 residual blocks, each containing 3 convolutional layers with small (15), medium (200), and large (500) kernel sizes to capture both local and global patterns in a spectrum. This design aligns with our tests, where we evaluated training efficiency and prediction accuracy using randomly selected kernel sizes (10 to 500 pixels) and model depths (8 to 72 layers). Our designed architecture represents an optimized balance of efficiency and accuracy.

For each residual block, the input spectrum \( x \) is processed through three convolutional layers initialized by the He Normal initializer \citep{he2015delving}. Batch normalization is applied after each convolutional layer to maintain network stability. The ReLU activation function \citep{xu2015empirical} is used following the first two batch normalization steps to introduce non-linearity, enabling the model to capture complex patterns. The output of each residual block \( H(x) \) is then added to the input \( x \) to form \( H(x) + x \), ensuring gradients remain within an appropriate range during training. After the residual blocks, the network flattens the output and passes it through fully connected layers, culminating in the classification of the spectrum.

 \subsection{Training}

A crucial aspect of training a neural network for a particular task is choosing the right loss function and optimization method. For classification tasks, "Cross Entropy" is commonly used, whereas "Mean Squared Error" is often preferred for regression tasks \citep{golik13_interspeech, 2023arXiv230407288M}. In the earlier version of the \code{FNet} model, we employed mean squared error since the task involved predicting the redshift of quasars \citep{Rastegarnia_2022}. In this work, "Cross Entropy" \citep[see e.g.,][and references therein]{2023arXiv230407288M} is used as the loss function for this classification problem, along with the \code{AdamW} optimizer to optimize the loss function.

\code{AdamW}, or \code{Adam with weight decay} \citep{2017arXiv171105101L}, is a modification of \code{Adam}, which combines the best properties of the AdaGrad \citep{JMLR:v12:duchi11a} and RMSProp \citep{Tieleman} algorithms to handle sparse gradients on noisy problems. \code{AdamW} retains the benefits of Adam’s adaptive learning rate and momentum while incorporating weight decay separately rather than adding it to the gradient as in \code{Adam}. This small but critical change helps in better regularization and often leads to better training outcomes. \code{AdamW} typically converges quickly and is effective in cases with large datasets and/or high-dimensional spaces. Experimental results demonstrate that \code{AdamW} leads to improved training loss and enhanced generalization compared to models trained with \code{Adam} \citep[see e.g.,][]{2017arXiv171105101L}.

We also performed a comparison by testing \code{FNet} on 40,000 spectra from the DR17Q catalog over 100 epochs (see Fig.~\ref{fig:comparisons}). The results clearly exhibit the advantage of using \code{AdamW}, which converges faster than other optimizers.

\begin{figure}
\centering
\includegraphics[width=1.0\hsize,clip]{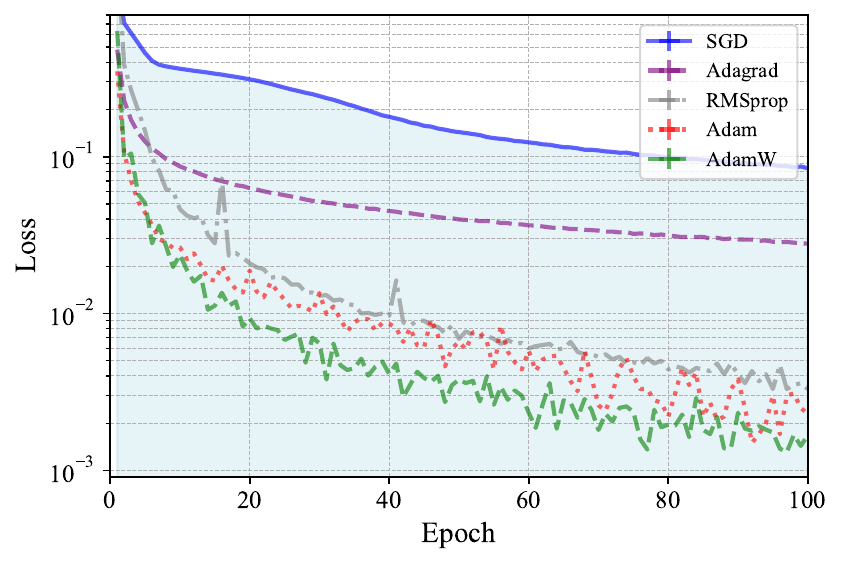}
\caption{The loss decay (on a logarithmic scale) of \code{FNet} for 40000 spectra of DR17Q catalog over 100 epochs for the \code{AdamW}, \code{Adam}, \code{SGD}, \code{AdaGrad}, and \code{RMSProp} optimizers is presented. The \code{AdamW} optimizer leads to a more rapid decline in the loss function, resulting in more accurate predictions.}  
\label{fig:comarisions}
\end{figure}

Optimizing a neural network model with the \code{AdamW} optimizer requires the right selection of hyperparameters. Experimentation with hyperparameter configurations is pivotal in identifying the optimal settings for achieving superior optimization and generalization outcomes. This includes fine-tuning the learning rate (lr), adjusting beta coefficients for gradient averaging, setting weight decay for regularization, and contemplating the utilization of AMSGrad \citep{2019arXiv190409237R}. 

In this work, specific parameter values were employed to obtain highly optimized results, including an initial learning rate of \(\rm lr_1=0.00024\), beta coefficients of (0.9, 0.999), weight decay of \(1\rm e{-6}\), and \code{amsgrad=False} to use the standard Adam update rule. Notably, the implementation of learning rate scheduling strategies, such as reducing the learning rate after designated epochs, can enhance convergence \citep[see][for more information about parameter selection and step learning rate]{Rastegarnia_2022}. Techniques like L2 penalty weight decay play a crucial role in mitigating overfitting, particularly in datasets characterized by a profusion of features \citep{lecun1998gradient, kingma2014adam}.

We utilized the \code{PyTorch} framework for training and testing our neural network model \citep{2019arXiv191201703P}. The source code for our model is available at: \href{https://github.com/AGNNet/FNet.git}{https://github.com/AGNNet/FNet.git}, and the dataset used can be accessed at: \href{https://www.kaggle.com/ywangscience/sdss-iii-iv}{https://www.kaggle.com/ywangscience/sdss-iii-iv}. Training of the \code{FNet} model was performed on an Nvidia Tesla V100 graphics card, with each training epoch taking approximately 90 minutes. The inference time for predicting the redshift of an SDSS spectrum is around 14 milliseconds.

\section{results}\label{sec:results}

As detailed in Sec.~\ref{sec:Data}, we used confident quasars, galaxies, stars, and BAL-quasars from the DR17Q catalog to train and test \code{FNet}.

We quantify the precision of the predictions using the metric of completeness \citep{2020ApJS..250....8L, 2018arXiv180809955B}, defined as the ratio of correctly predicted spectra to the total number of spectra in each class. This metric is assessed on the {test} sample.

\subsection{Imbalance in Data}\label{sec:imbalance}

Highly imbalanced data presents a significant challenge, as most networks tend to prioritize the majority class. In extreme cases where the minority class constitutes less than 1\%, it may be completely disregarded. Addressing class imbalance in deep learning is an active area of research, and finding effective approaches for specific tasks and imbalance scenarios is challenging \citep{2022arXiv221004675H}. However, traditional methods such as data sampling and cost-sensitive learning have shown applicability in deep learning contexts in recent years \citep{Johnson2019SurveyOD}.

In our sample, most spectra belong to quasars (approximately 390,000) and stars (approximately 230,000), while fewer spectra are attributed to BAL-quasars (approximately 26,000) and galaxies (approximately 30,000). This creates a moderate imbalance, with only 9\% of the data representing the less common classes. Testing the \code{FNet} model on this imbalanced data yielded expected results: galaxy completeness was 86.9\%, BAL-quasar completeness was 78.8\%, star completeness was 98.8\%, and quasar completeness was 98.3\%. As anticipated, training the network with this imbalanced sample prioritized quasars and stars while neglecting BAL-quasars and galaxies. We attempted oversampling the minority classes and using weighted cross-entropy to address the imbalance, but these methods did not significantly improve the results.

Fortunately, the number of minority class spectra (approximately 30,000 for each class) in our sample is sufficient to establish a confident training set. To mitigate data imbalance, we equalized the number of spectra for each class to \( N = 26,667 \), the total number of BAL quasars, which is the smallest class.

\subsection{Problem of BAL-quasar identification}

It is widely acknowledged that detecting quasars with broad absorption line (BAL) features \citep{1981ARA&A..19...41W, 1991ApJ...373...23W} is more complex than detecting emission lines at the human-expert level. Quasars with BAL features display a distinct absorption trough blueward of the emission lines, which must be sufficiently separated from the line center and possess adequate width. The balnicity index (BI), an automated measurement of the strength of the BAL feature in the CIV line, annotated in the SDSS survey as \code{BI\_CIV}, quantifies this characteristic. The primary challenge in determining whether a spectrum exhibits a BAL feature, whether through visual inspection or automated methods, lies in accurately assessing the level of the unabsorbed continuum.

For example, in this context, \code{QuasarNET} \citep{2018arXiv180809955B} addresses the problem of identifying a CIV emission line with broad absorptions by treating it as equivalent to detecting any other emission line. To achieve this, they incorporate a \code{BAL\_CIV} line into the list of lines that the network is trained to recognize. \code{QuasarNET}, which consists of a network with four convolutional layers followed by a final fully-connected layer, includes "six line finders" to detect Ly$\alpha$ (121.6 nm), CIV (154.9 nm), CIII (190.9 nm), MgII (279.6 nm), H$\beta$ (486.2 nm), and H$\alpha$ (656.3 nm) emission lines. Additionally, it is enhanced with a seventh ``line finder'' unit specifically designed to identify the CIV line with a broad absorption feature.

On the other hand, \code{FNet} utilizes a ResNet structure with 24 convolutional layers and kernel sizes of 500, 200, and 15. This design enables \code{FNet} to recognize both ``global'' and ``local'' patterns in spectra through a self-learning procedure, independently discovering hidden features and training on all spectra uniformly. The presence of the larger kernels, particularly the size 500, effectively bypasses complexities in detecting BAL features and eliminates the need to incorporate external information, such as the \code{BAL\_CIV} threshold, for spectral classification. 

We tested \code{FNet} using only smaller kernel sizes of 100 and 10, and 200 and 17, instead of 500, 200, and 15. The completeness decreased with the smaller kernels, especially for BAL-quasars, suggesting that some broad features were overlooked by the network.

\subsection{training results}\label{sec:result}

Eighty percent of the spectra were allocated to the training set, while 10\% were reserved for validation using k-fold cross-validation \citep{refaeilzadeh2009cross}. The remaining spectra were designated for the test set, ensuring they were not used to train the CNN.

\begin{figure}
\centering
\includegraphics[width=1.0\hsize,clip]{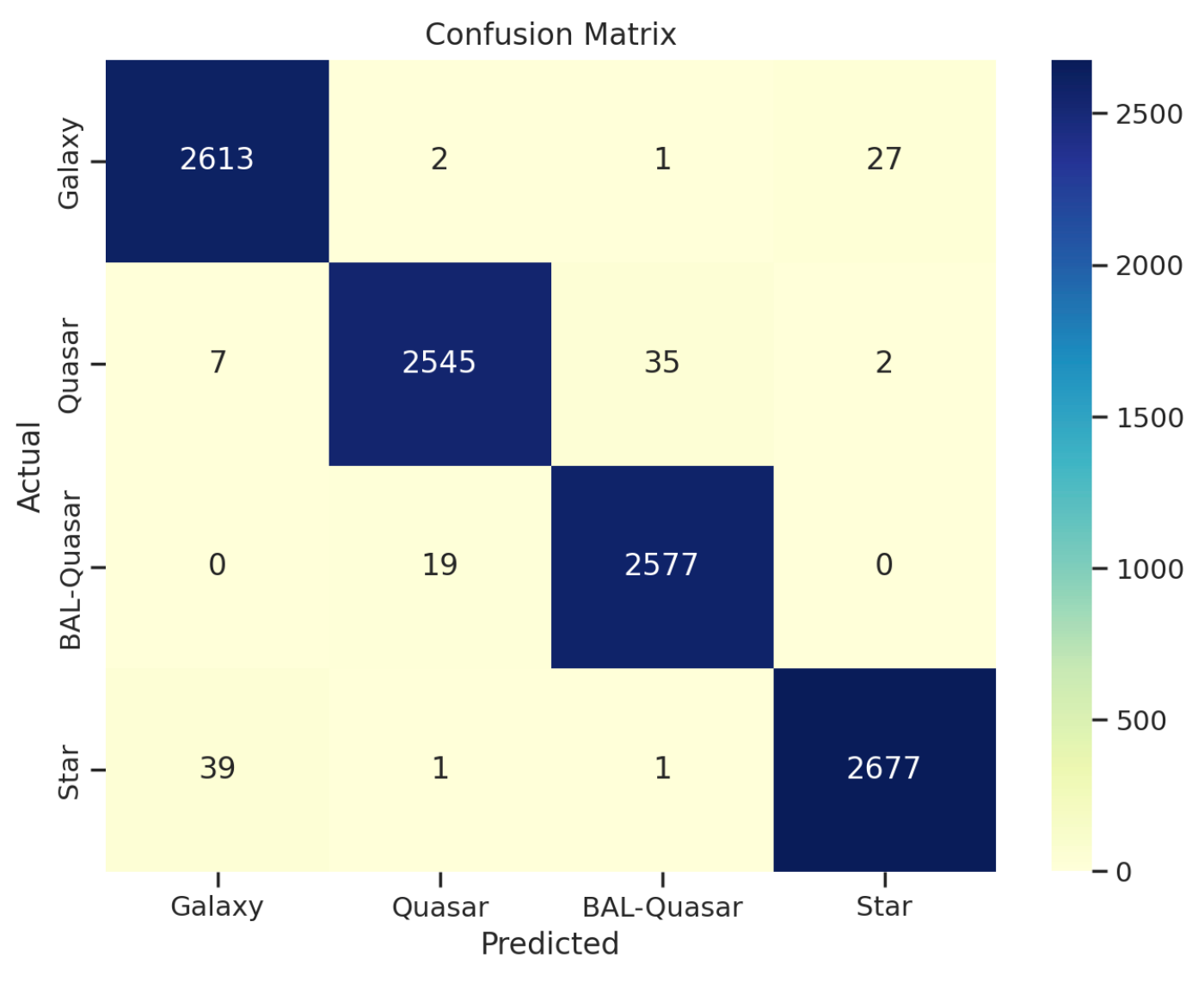}
\caption{The confusion matrix, evaluating the FNet classification. The classification completeness achieved by \code{FNet} is 99.2\% for galaxies, 98.6\% for quasars, 99.0\% for BAL-quasars, and 98.9\% for stars. }  
\label{fig:confusion}
\end{figure}

\begin{figure*}
\centering
\textbf{A}\includegraphics[width=0.8\hsize,clip]{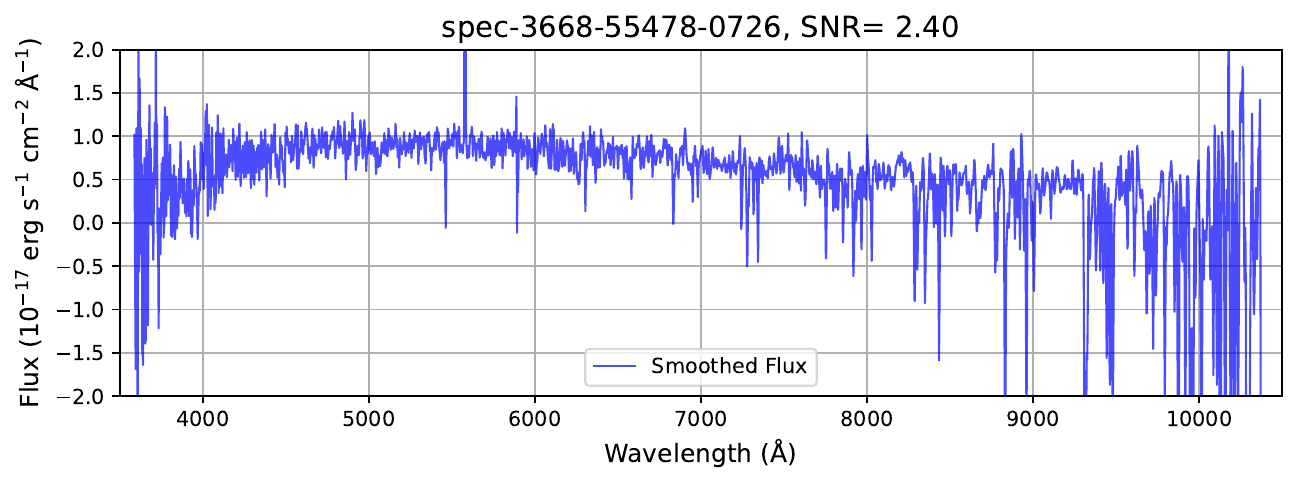}
\textbf{B}\includegraphics[width=0.8\hsize,clip]{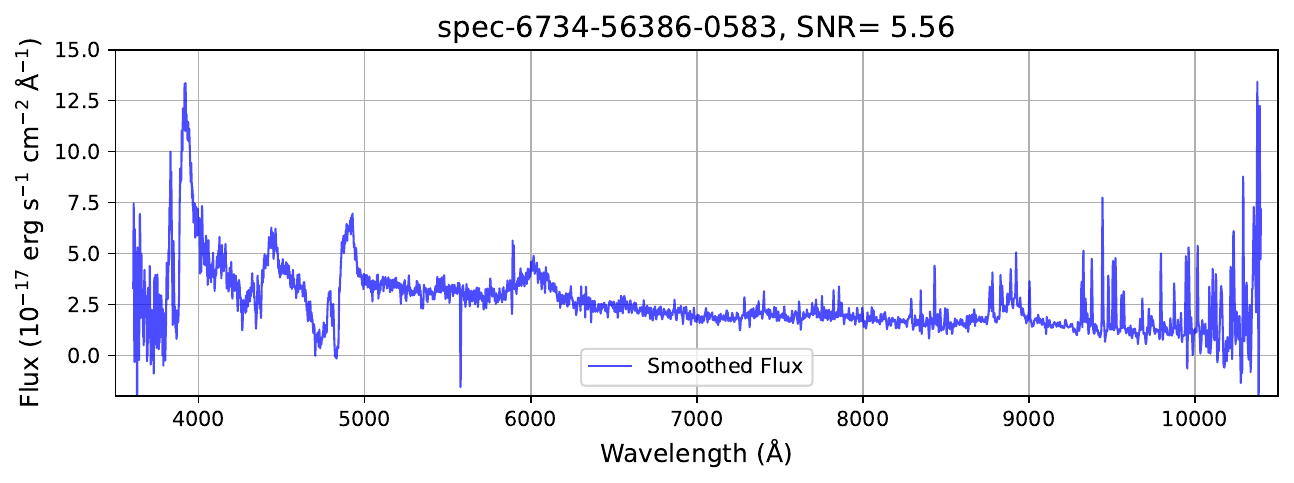}
\textbf{C}\includegraphics[width=0.8\hsize,clip]{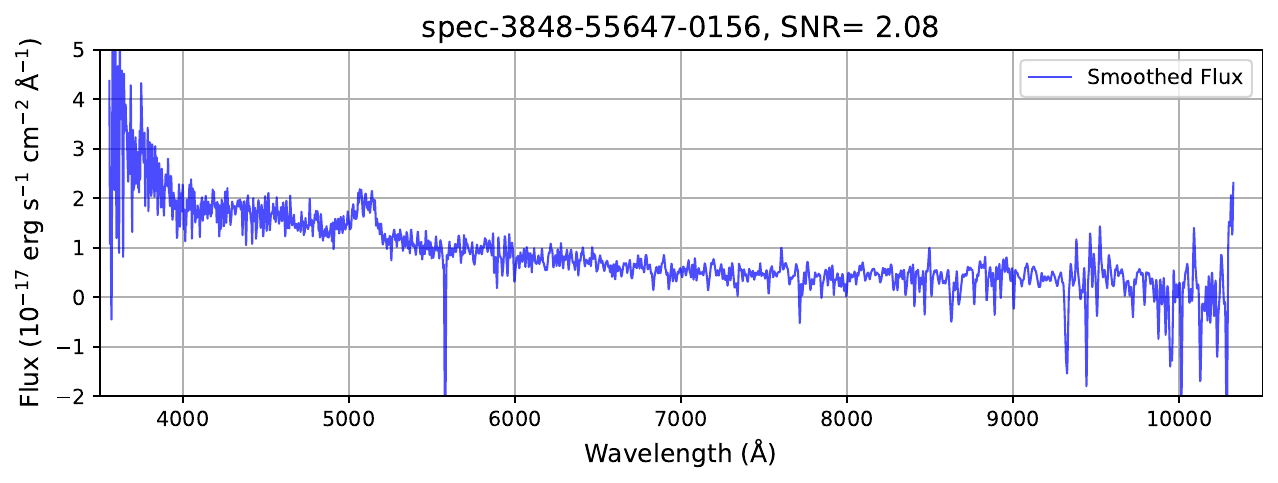}
\textbf{D}\includegraphics[width=0.8\hsize,clip]{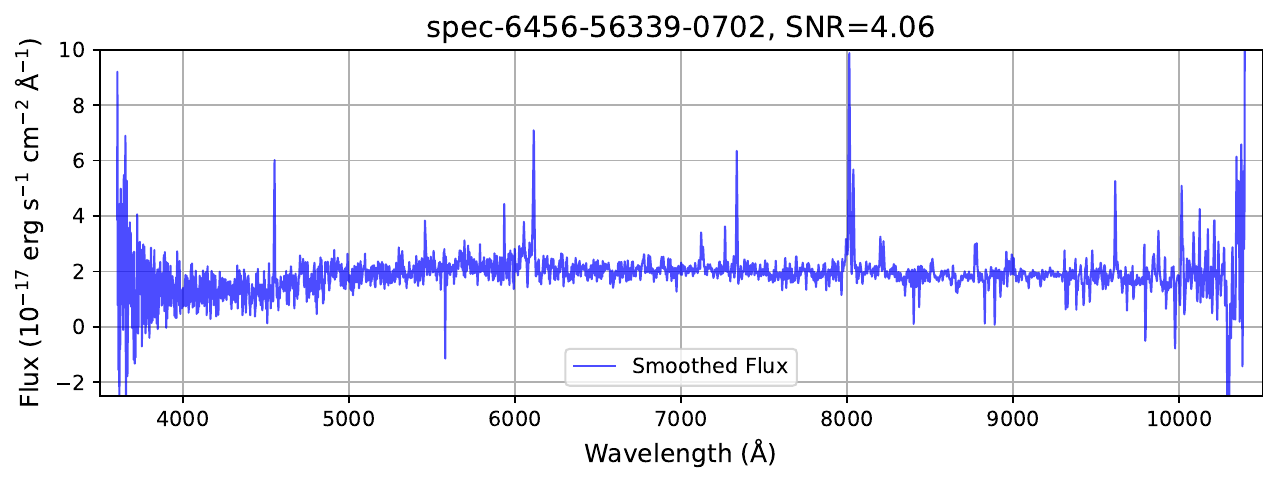}
\caption{{Samples of true classification by \code{FNet}. (A) Star, (B) BAL-quasar, (C) Quasar, and (D) Galaxy. Although the signal-to-noise ratio (SNR) is relatively low in these samples, the \code{DELTA\_CHI2} values are high.
}}  
\label{fig:DR14Q-JJ}
\end{figure*}

\begin{figure*}
\centering
\textbf{A}\includegraphics[width=0.45\hsize,clip]{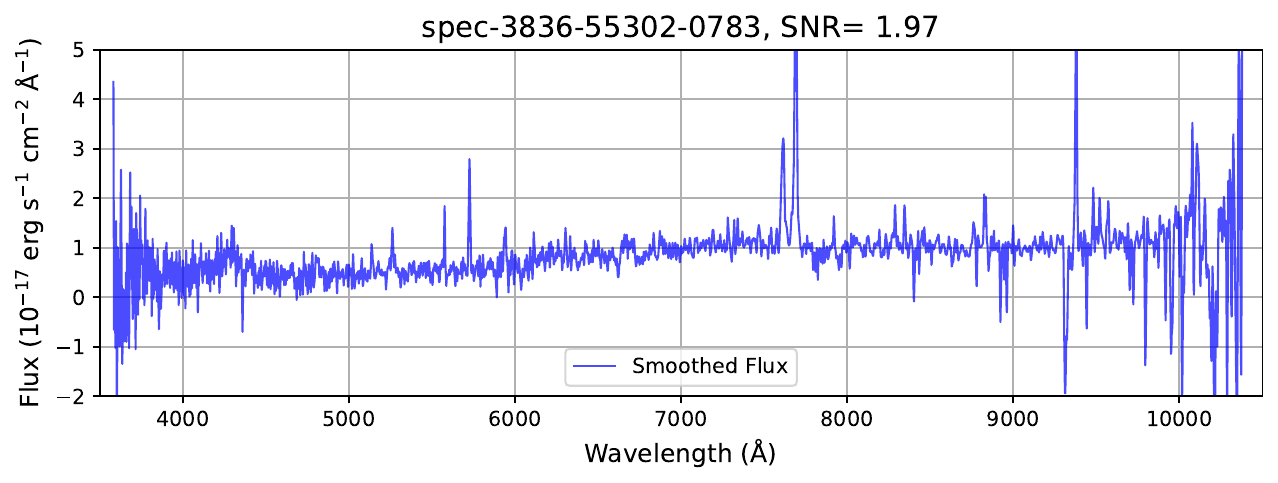}
\textbf{B}\includegraphics[width=0.45\hsize,clip]{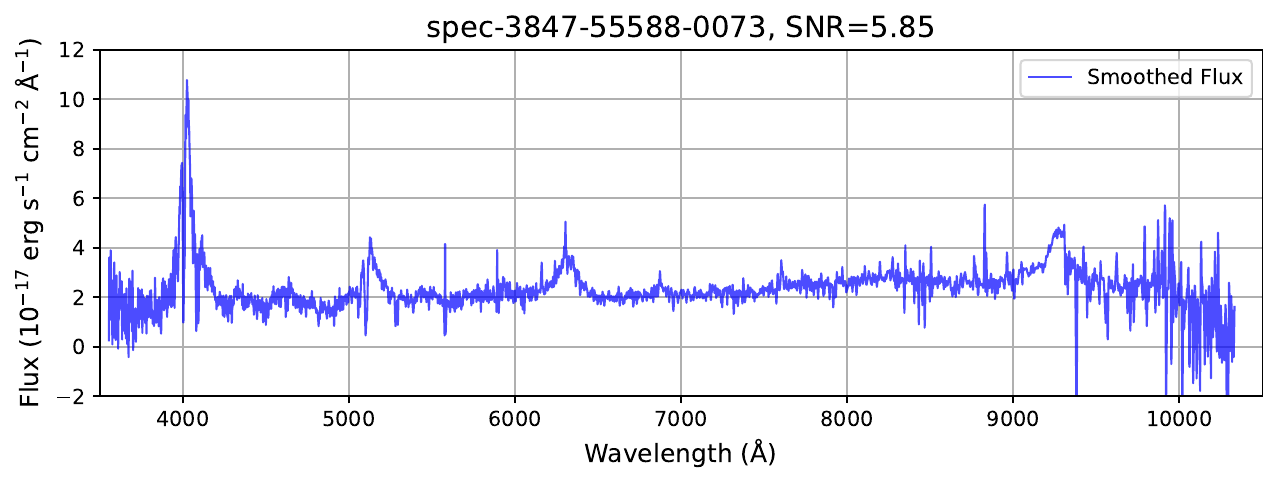}
\textbf{C}\includegraphics[width=0.45\hsize,clip]{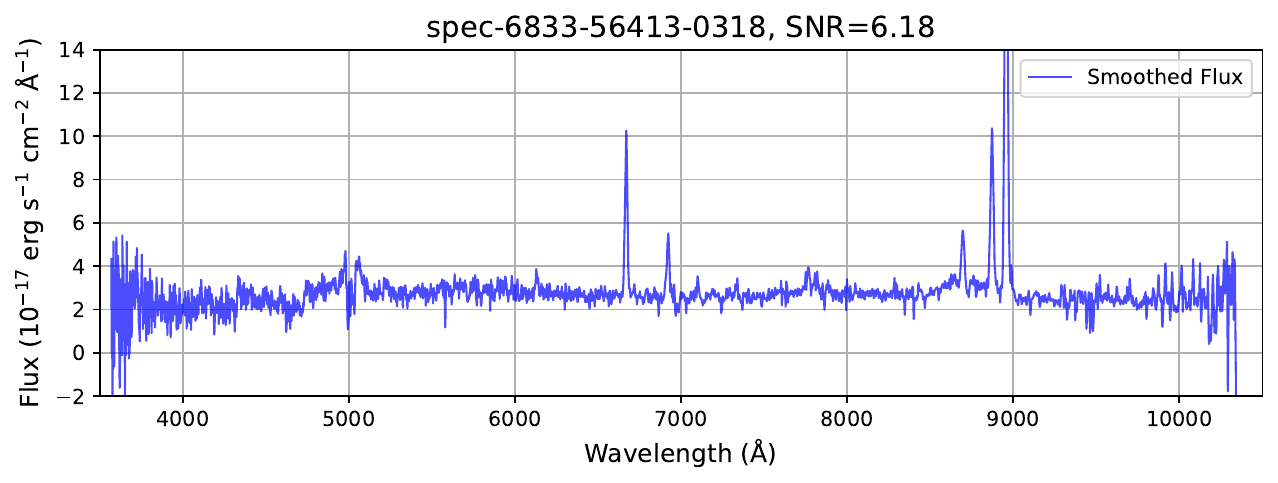}
\textbf{D}\includegraphics[width=0.45\hsize,clip]{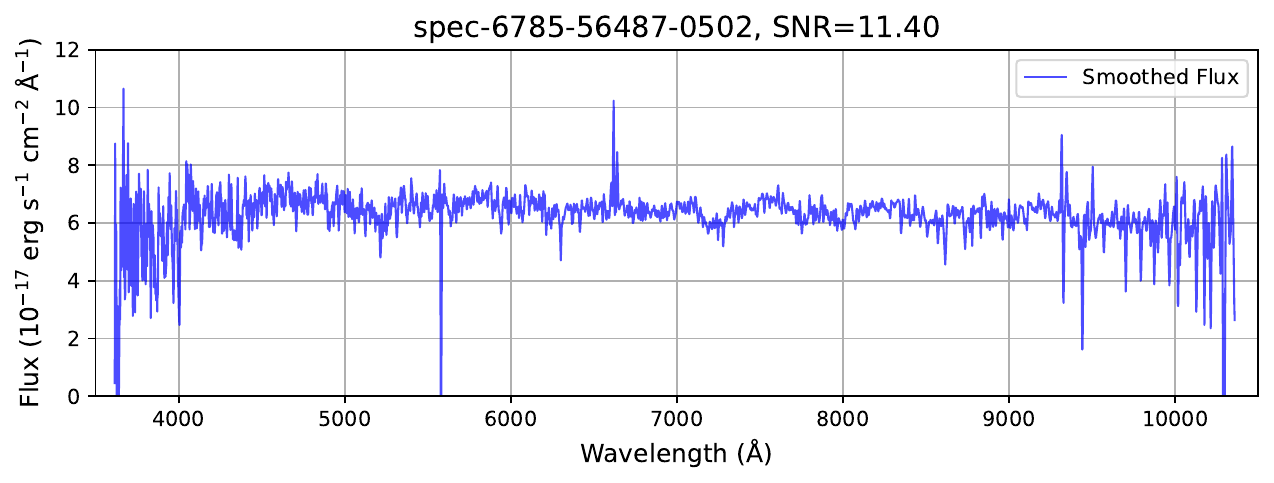}
\textbf{E}\includegraphics[width=0.45\hsize,clip]{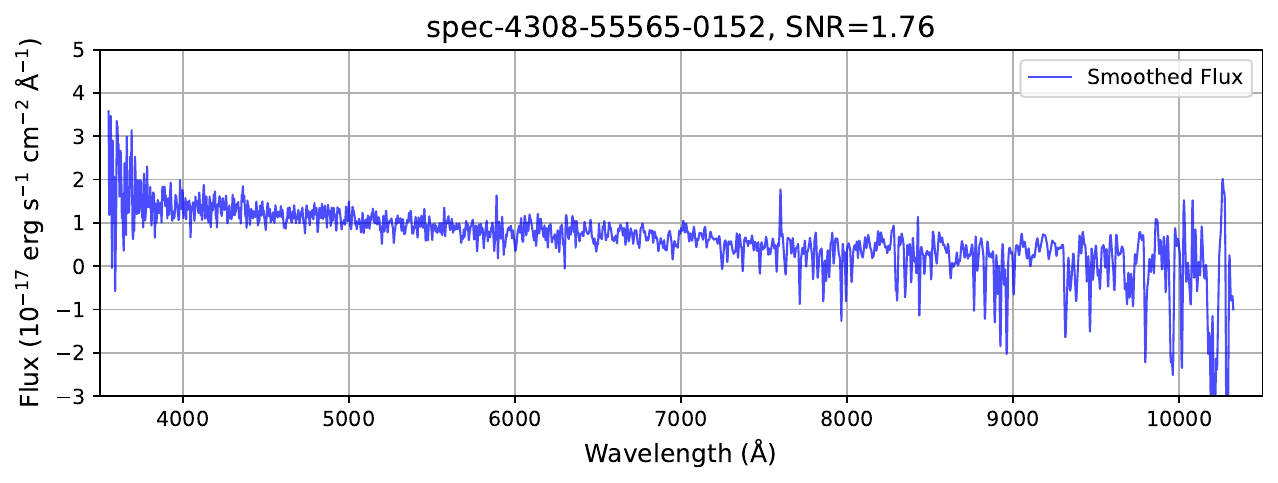}
\textbf{F}\includegraphics[width=0.45\hsize,clip]{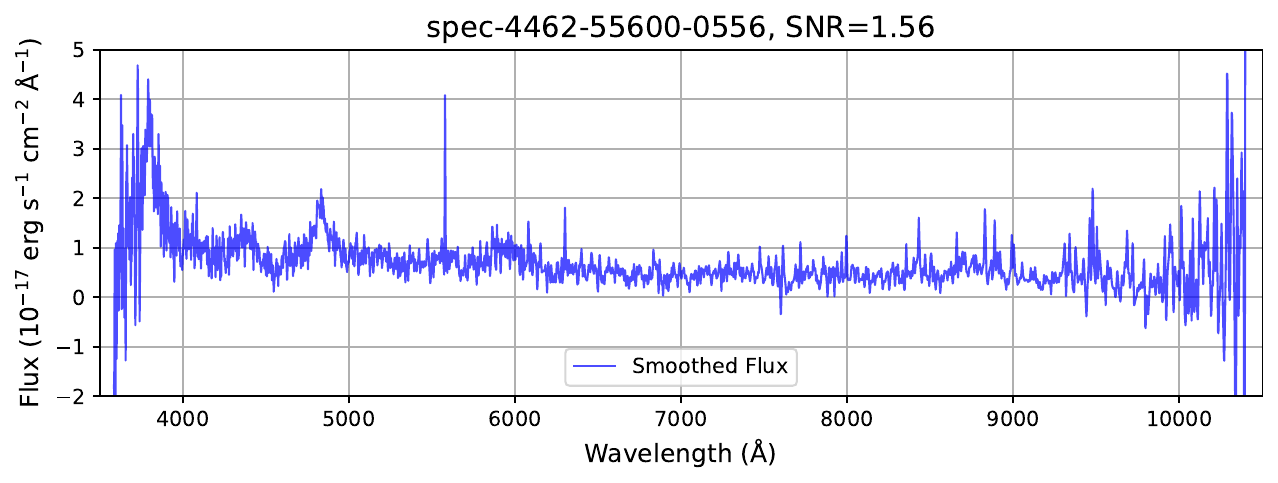}
\textbf{G}\includegraphics[width=0.45\hsize,clip]{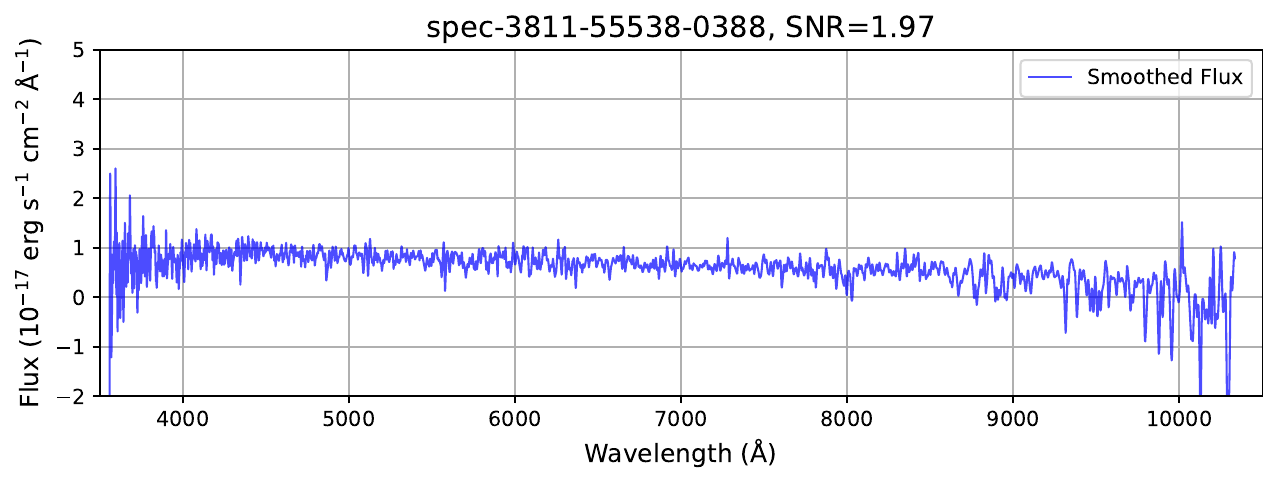}
\textbf{H}\includegraphics[width=0.45\hsize,clip]{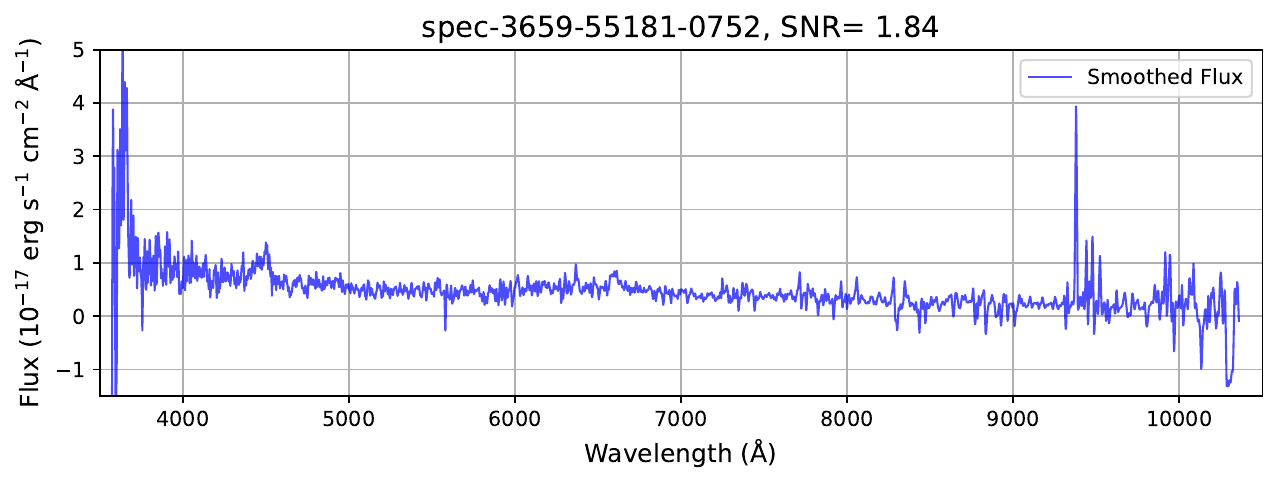}
\caption{{Samples of incorrect classifications by \code{FNet}. (A) Quasar predicted as Galaxy, (B) Quasar predicted as BAL-quasar, (C) Galaxy predicted as Quasar, (D) Galaxy predicted as BAL-quasar, (E) Galaxy predicted as Star, (F) BAL-quasar predicted as Quasar, (G) Star predicted as Galaxy, and (H) Star predicted as Quasar. The signal-to-noise ratio (SNR) for the majority of these sources is relatively low, and the \code{DELTA\_CHI2} values are regular. Notably, more than 60\% of quasars incorrectly predicted as stars, despite having \code{CLASS\_PERSON = 3}, are also annotated as \code{AUTOCLASS\_DR14Q = STAR}. Similarly, more than 60\% of stars incorrectly predicted as galaxies, despite having \code{CLASS\_PERSON = 1}, are also annotated as \code{AUTOCLASS\_DR14Q = GALAXY}. Additionally, almost all stars predicted as quasars are annotated as \code{AUTOCLASS\_DR14Q = QUASAR} despite having \code{CLASS\_PERSON = 1}. Therefore, the result for completeness of \code{FNet}'s classification for these sources remains uncertain, and it is not evident whether \code{FNet}'s classifications are actually incorrect.  Study the reliability of this issue requires more careful research, which will be addressed in future work.
}}  
\label{fig:DR14Q-JK}
\end{figure*}

After exclusively using the training/test set {primarily sourced from our criteria introduced in Sec.~\ref{sec:Data}} and excluding those labeled with \code{AUTOCLASS\_DR14Q = VI}, consistent with the approach employed in \code{QuasarNET} \citep{2018arXiv180809955B}, the classification completeness achieved by \code{FNet} improved to 99.00\% $\pm$ 0.20 for galaxies, 98.50\% $\pm$ 0.30 for quasars, 99.00\% $\pm$ 0.18 for BAL-quasars, and 98.80\% $\pm$ 0.20 for stars. The uncertainty in completeness percentages reflects tests conducted over multiple data batches. This method of evaluating a model is particularly useful for ensuring robustness and understanding the consistency of the model's performance, with validity ensured by choosing large batches that include randomly selected test samples of all classes. The confusion matrix evaluating \code{FNet}'s classification is depicted in Fig.~\ref{fig:confusion}. Figure~\ref{fig:DR14Q-JJ} and Figure~\ref{fig:DR14Q-JK} show examples of both successes and failures of the final test. {In order to reduce noise in the data and make underlying spectral features, such as emission or absorption lines, more pronounced, we used Gaussian smoothing on all spectra in the above figures only for presenting purpose.}

Indeed, when training on imbalanced data (as illustrated in Sec.~\ref{sec:imbalance}), the high number of misclassified rare sources indicates that the network tends to confuse rare sources with dominant ones, reducing the completeness for dominant sources. Conversely, using balanced data yields more accurate results for rare sources. This improvement in completeness for rare sources subsequently increases the overall number of correctly predicted sources, as there are fewer instances of rare sources being confused with dominant ones.

\begin{figure}
\centering
\includegraphics[width=1.0\hsize,clip]{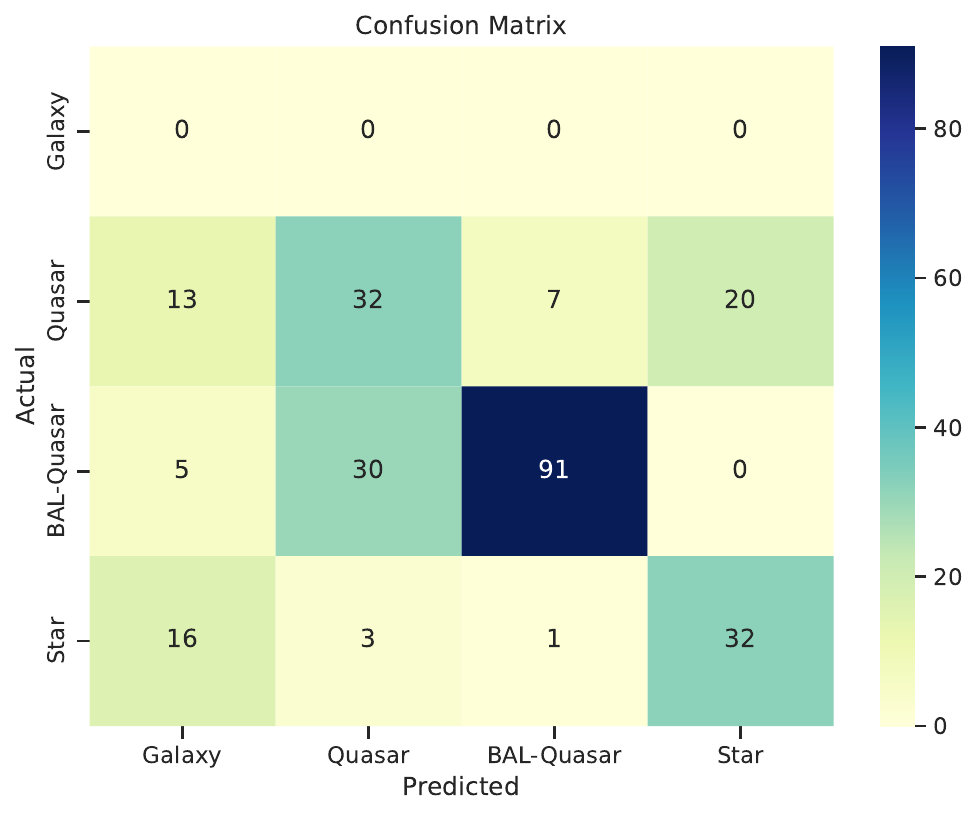}
\caption{{The confusion matrix of the \code{FNet} classification for spectra with low redshift confidence (\code{Z\_CONF} $\neq$ 3), less secure BAL probabilities (0.5 $\leq$ \code{BAL\_PROB} $<$ 0.9), and ambiguously classified spectra as quasars (\code{IS\_QSO\_FINAL} $\neq$ 1). The classification completeness achieved by \code{FNet} is 45\% for quasars, 72\% for BAL-quasars, and 62\% for stars. For galaxies, since the number of sources with \code{Z\_CONF} $\neq$ 3 is negligible, we do not consider them for this test.}}  
\label{fig:confusion1}
\end{figure}

{\subsection{Training result for less confident spectra and potential application to other datasets} }

In Sec.~\ref{sec:Data}, we used only data with the highest confidence redshifts (\code{Z\_CONF} = 3), the most secure BAL probabilities (\code{BAL\_PROB} $\geq$ 0.9), and no ambiguity about quasar classification (\code{IS\_QSO\_FINAL} = 1). Relaxing these conditions to include sources with low redshift confidence (\code{Z\_CONF} $\neq$ 3), less secure BAL probabilities (0.5 $\leq$ \code{BAL\_PROB} $<$ 0.9), and ambiguously classified quasars (\code{IS\_QSO\_FINAL} $\neq$ 1) adds an extra 2\% of spectra to the main data for the training-test procedure. These spectra are mostly related to sources with too low SNRs to be confidently classified by human experts \citep[see also][]{2018arXiv180809955B, 2020ApJS..250....8L}.

We performed training and testing on the dataset with ambiguous data included and observed a slight decrease in completeness compared to using only clean data. Specifically, \code{FNet} achieves a completeness of 98.30\% $\pm$ 0.40 for galaxies, 98.00\% $\pm$ 0.40 for quasars, 98.50\% $\pm$ 0.40 for BAL-quasars, and 98.20\% $\pm$ 0.30 for stars. These results remain acceptable for scientific use.

For example, when testing \code{QuasarNet} on the sample of spectra without confident annotations (\code{Z\_CONF} < 3), which represents about 2\% of the spectra \citep{2018arXiv180809955B}, it classifies 20\% of the spectra as quasars with reliable redshifts. In comparison, \code{FNet} achieves a completeness of 45\% for quasars, 72\% for BAL-quasars, and 62\% for stars. For galaxies, the number of sources with \code{Z\_CONF} $\neq$ 3 is negligible, so we do not consider them for this test. Fig.~\ref{fig:confusion1} shows the confusion matrix for this 2\% of ambiguous spectra.

In general, \code{FNet} learns to classify spectra not only by recognizing prominent emission and absorption lines but also by identifying subtle patterns that may not be immediately apparent \citep{Rastegarnia_2022}. Consequently, even after incorporating the 2\% of train/test sets with less confident classifications, \code{FNet} still achieves high classification completeness. This robustness suggests that \code{FNet} has significant potential for application to other datasets. However, careful data preprocessing is essential when applying \code{FNet} to datasets other than SDSS-IV.\\

\section{Conclusions}\label{sec:conclusions} 

Accurate spectral classification in astrophysics requires reliable methods. With the increasing volume of data from current and forthcoming surveys, relying solely on visual inspection is impractical. Research into statistical approaches for analyzing spectra within their cosmological context \citep{2023MNRAS.518.5904Y} has shown that convolutional neural networks (CNNs) outperform other methods in classifying observed spectra. Therefore, developing automated methods that achieve human-expert accuracy has become essential.

In this study, we adapted and modified the previously introduced network, \code{FNet}, a 1-dimensional CNN with a ResNet architecture, to classify the spectra of quasars, galaxies, stars, and BAL-quasars in the SDSS-IV catalog from the DR17 superset of eBOSS. This network was previously successful in predicting the redshift of quasars in DR16 of eBOSS, achieving a velocity difference accuracy of 97.0\% for $|\Delta \nu| < 6000 \ \rm km/s$, 98.0\% for $|\Delta \nu| < 12000 \ \rm km/s$, and 98.9\% for $|\Delta \nu| < 30000 \ \rm km/s$ when compared to visually inspected redshifts \citep{Rastegarnia_2022}.

As \code{FNet} does not rely on identifying specific lines, a simple modification enabled our current network to classify all SDSS spectra without external information about emission/absorption lines. This modification involves changing the last output layer of \code{FNet} from a single value (redshift) to multiple values (probabilities of all classes) and adjusting the loss function from mean squared error (MSE) to cross-entropy.

From a total of 1,367,288 spectra of quasars, stars, and galaxies in the SDSS-IV quasar catalog from DR17 of eBOSS, we selected 639,906 spectra (Stars: 229,647; Galaxies: 28,688; Quasars: 359,609; BAL Quasars: 21,962) for the classification training set. This subset represents the highest confidence level of source selection in this catalog.

Utilizing its convolutional layers and the ResNet structure, \code{FNet} autonomously discerns both ``local'' and ``global'' patterns within the spectra. The CNN employs convolutional layers with kernel sizes of 500, 200, and 15 to detect various patterns in the flux of quasars. The flux is then processed through 24 residual blocks: the first 21 blocks have a channel size of 32, followed by three blocks with channel sizes of 64, 32, and 16, respectively. This self-learning capability is crucial for classifying spectra from diverse astrophysical objects.

It is worth noting that detecting quasars with broad absorption lines (BALs), as in the case of BAL quasars, is slightly more challenging than detecting emission lines. Unlike \code{QuasarNET}, which addresses this by adding the BAL CIV line to the list of identifiable lines to find a CIV emission line with broad absorptions, this step is unnecessary in \code{FNet} due to its self-learning feature acquisition process.

\code{FNet} achieves a completeness of 99.00\% $\pm$ 0.20 for galaxies, 98.50\% $\pm$ 0.30 for quasars, 99.00\% $\pm$ 0.18 for BAL quasars, and 98.80\% $\pm$ 0.20 for stars. These results are comparable to those obtained using \code{QuasarNET} \citep{2018arXiv180809955B}, a standard CNN used in the SDSS catalog for redshift measurement and classification \citep{2020ApJS..250....8L}.

\section*{Acknowledgment}

We acknowledge the support from Dr. Jie Jiang and Prof. Yifu Cai from USTC for providing the Nvidia DGX Station. {We would like to thank the anonymous referee for their comments, which have enhanced the scientific credibility of the paper, leading to more comprehensive discussions and clearer explanations. R. Moradi acknowledges support from the Institute of High Energy Physics, Chinese Academy of Sciences (E32984U810).}

\section*{Data Availability}

The catalog underlying this article is available in The Sloan Digital Sky Survey Quasar catalog: sixteenth data release (DR16Q), at \href{https://www.sdss.org/dr16/algorithms/qso_catalog/}{https://www.sdss.org/dr16/algorithms/qsocatalog/}. The code can be found in the public domain: \href{https://github.com/AGNNet/FNet.git}{https://github.com/AGNNet/FNet.git} and the prepared data set can be found in \href{https://www.kaggle.com/ywangscience/sdss-iii-iv}{https://www.kaggle.com/ywangscience/sdss-iii-iv}.

\bibliographystyle{mnras}

\end{document}